\documentclass[12pt]{article}

\usepackage{amssymb,amscd,array,latexsym}

\def\Journal#1#2#3#4{{#1} {\bf #2}, #3 (#4)}

\def\NPB{{\em Nucl. Phys.} B}

\def\PRD{{\em Phys. Rev.} D}

\def\be{\begin{equation}}
\def\ee{\end{equation}}
\def\bea{\begin{eqnarray}}
\def\eea{\end{eqnarray}}
\begin{document}
\begin{titlepage}
\begin{center}
{\bf {\Large {SOME ASPECTS OF QUANTUM GRAVITY ~\\ IN THE CAUSAL
 APPROACH \\}}}

\end{center}
\vskip 1.0truecm
\centerline{N. GRILLO
\footnote{e-mail: grillo@physik.unizh.ch}}
\vskip 5mm
\centerline{\it{Institut f\"ur Theoretische Physik,
 Universit\"at Z\"urich}}
 \centerline{\it{Winterthurerstrasse 190,
 CH-8057 Z\"urich, Switzerland}  }
\vskip 2cm
\bigskip \nopagebreak
\begin{abstract}
\noindent

We describe the construction of
quantum gravity, {\em i.e.} of a theory of self-interacting massless
spin-2 quantum gauge fields, the `gravitons', on flat space-time, in the framework of causal perturbation theory.

\end{abstract}
\vskip 7cm
\sl{Talk given at the 4th Workshop on Quantum Field Theory under the
 Influence of External Condition, Leipzig, Germany, 14-18 Sep 1998}

\end{titlepage}
 
\section{Introduction}

The central aspect of this work is the construction of the $S$-matrix
by means of causality in the quantum field theoretical (QFT) framework.
This idea goes back to St\"uckelberg, Bogoliubov and Shirkov and the
 program
was carried out correctly by Epstein and Glaser~\cite{eg} for scalar
 field theory and
subsequently applied to QED by Scharf~\cite{scha1} and to non-abelian
 gauge theories
by D\"utsch {\it et al.}~\cite{du1,aste}. We now apply this scheme
to `quantum gravity' (QG),
{\em i.e.} a QFT of self-interacting massless spin-2 quantum gauge fields
 on
flat space-time.
For this purpose, two main tools will be used: the Epstein-Glaser inductive construction of the perturbation series for the $S$-matrix with the related causal renormalization scheme and  perturbative quantum gauge invariance~\cite{du1,aste}.
The first method provides an elegant way of dealing with the
UV problem of QG and the second one ensures gauge invariance at the quantum level,
formulated by means of the `gauge charge' $Q$, in each order of
 perturbation theory.
A detailed exposition will be found in forthcoming papers by the author.

\section{Causal Perturbation Theory}

We consider the $S$-matrix, a formal power series in the coupling constant,
as a sum of smeared operator-valued distributions of the following form~\cite{eg}:
\begin{equation}
S(g)={\mathbf 1}+\sum_{n=1}^{\infty} \int d^{4}x_{1}\ldots d^{4}x_{n}\, T_{n}(x_{1},\ldots,
x_{n})\,g(x_{1})\cdot\ldots\cdot g(x_{n}),
\label{eq:1}
\end{equation}
where $g$ is a Schwartz test function which plays the r{\^o}le of
adiabatic switching of the interaction and provides a natural infrared
 cut off
in the long-range part of the interaction.
The $T_{n}$, n-point operator-valued distributions,
are well-defined  `renormalized time-ordered products' and can be
expressed in term of Wick monomials of free fields. They are constructed
inductively from the first order $T_{1}(x)$, which defines the theory, by
 means of Poincar\'e covariance and
causality; the latter, if correctly incorporated, leads directly to the
finite perturbation series for the $S$-matrix. The construction of
$T_{n}$ requires some care: if it were simply given by the usual
 time-orderig
$T_{n}(x_{1},\ldots,x_{n})=T\{T_{1}(x_{1})\ldots T_{1}(x_{n})\}$, then UV-divergences would appear. If the arguments
$x_{1},\ldots ,x_{n}$ are all time-ordered, {\em i.e.} if we have $x_{1}^{0}>x_{2}^{0}>
\ldots > x_{n}^{0}$, then $T_{n}$ is rigorously given by $T_{n}(x_{1},\ldots,x_{n})=T_{1}(x_{1})\ldots T_{1}(x_{n})$; since $T_{n}$
is totally symmetric in $x_{1},\ldots ,x_{n}$, we obtain $T_{n}$
everywhere except for the complete diagonal $\Delta_{n}=\{x_{1}=x_{2}=
\ldots =x_{n}\}$. After performing Wick expansion of $T_{n}$, we can
 extend
the c-number distributions from $R^{4n}\setminus \Delta_{n}$ to
 $ R^{4n}$, so
that we obtain
\begin{equation}
\begin{array}{rcl}
 T_{n}(x_{1},\ldots,x_{n})&=&\sum_{k} :{\mathcal{O}}_{k}(x_{1},
\ldots , x_{n}):
             t_{n}^{k}(x_{1},\ldots , x_{n}),\\ \\
t_{n}^{k}(x_{1},\ldots , x_{n})&=& \tilde{t}_{n}^{k}(x_{1},
\ldots , x_{n})+
\hbox{sum of free local normalization terms}
\end{array}
\label{eq:2}
\end{equation}
where $:{\mathcal{O}}_{k}(x_{1},\ldots , x_{n}):$ represents a normally
 ordered 
product of free field operators and $t_{n}^{k}(x_{1},\ldots , x_{n})$ a
well-defined c-number distribution which is not unique: it is 
ambiguous up to distributions with local support $\Delta_{n}$ which
depend on the power counting degree of the distribution.
This normalization freedom has to be restricted by further physical
conditions. In momentum space, $\tilde{t}_{n}^{k}(x_{1},\ldots , x_{n})$
 is
 best obtained by means of dispersion-like integrals~\cite{scha1} which correspond
to the splitting of causal distributions into retarded and advanced
 parts
with respect to the last argument $x_{n}$ (see Sec.~\ref{sec:causal}).

\section{Quantization of Gravity}

For the causal construction we need
the equation of motion of the free graviton
after fixing the gauge, the commutation relation between free fields
at different space-time points and the first-order graviton self-coupling $T_{1}^{h}(x)$.
Since we are interested in a quantum theory of Einstein's general
 relativity,
we therefore start from the Hilbert-Einstein Lagrangian density
${\mathcal{L}}_{\scriptscriptstyle HE}$ written in terms of the 
Goldberg
variable $\tilde{g}^{\mu\nu}=\sqrt{-g}g^{\mu\nu}$ and by expanding it
 into
a power series in the coupling constant $\kappa =32\pi G $, by
 introducing
the `graviton' field $h^{\mu\nu}$ defined through
$\kappa h^{\mu\nu}=\tilde{g}^{\mu\nu}-\eta^{\mu\nu}$, where
 $\eta^{\mu\nu}$ is the flat space-time metric tensor:
\begin{equation}
{\mathcal{L}}_{\scriptscriptstyle HE}={\frac{- 2}{\kappa ^ 2}}\sqrt{-g}
 R =
\sum_{j=0}^{\infty}\kappa^{j} {\mathcal{L}}_{\scriptscriptstyle HE}^{\scriptscriptstyle (j)};
\label{eq:3}
\end{equation}
and ${\mathcal{L}}_{ \scriptscriptstyle HE}^{\scriptscriptstyle (j)}$
 represents  an `interaction' involving $j+2$ gravitons. From this
 formulation of general relativity we extract the ingredients
for the perturbative construction of causal QG. We stress however the
fact that we consider the classical Lagrangian density Eq.~(\ref{eq:3})
only as a `source' of information about the fields, the couplings and the
gauge which we work with: causal perturbation theory doesn't rely on any
quantum Lagrangian.
By considering the Euler-Lagrange variation of ${\mathcal{L}}_{\scriptscriptstyle HE}^{\scriptscriptstyle (0)}$
from  Eq.~(\ref{eq:3}) in the Hilbert-gauge $h^{\alpha\beta}_{,\beta}=0$
 we
obtain the equation of motion for the free graviton field
$\Box h^{\alpha\beta}(x)=0$, and quantize it covariantly
by imposing the commutation rule
\begin{equation}
\left[ h^{\alpha\beta}(x),h^{\mu\nu}(y) \right]=-i
{\underbrace{\frac{1}{2}\left(  \eta^{\alpha\mu}\eta^{\beta\nu}+   \eta^{\alpha\nu}\eta^{\beta\mu}-\eta^{\alpha\beta}\eta^{\mu\nu}\right) }_{\displaystyle b^{\alpha\beta\mu\nu}}}
 D_{0}(x-y),
\label{eq:4}
\end{equation}
where $D_{0}(x)$ is the Jordan-Pauli causal distribution.
The first order coupling among gravitons, being linear in the coupling
 constant $\kappa$,
can be derived from Eq.~(\ref{eq:3}) by taking the normally ordered

 product of ${\mathcal{L}}_{\scriptscriptstyle HE}^
{\scriptscriptstyle(1)}$ :
\begin{equation}
T_{1}^{h}(x)=i\kappa :{\mathcal{L}}_{\scriptscriptstyle HE}^{\scriptscriptstyle(1)}(x):=i
\frac{\kappa}{2}\left\{
:h^{\alpha\beta}(x)h^{\rho\sigma}(x)_{,\alpha}
h_{\rho\sigma}(x)_{,\beta}:\ldots\right\}.
\label{eq:5}
\end{equation}

\section{Perturbative Quantum Gauge Invariance}

The classical gauge transformations $h^{\alpha\beta} \to h^{\alpha\beta}
+u^{\alpha ,\beta}+u^{\beta ,\alpha} - 
\eta^{\alpha\beta} u^{\sigma}_{,\sigma}$
can be quantum mechanically implemented in the following way by means
 of the `gauge charge' $Q$:
\begin {equation}
\begin{array}{rcl}
h'^{\alpha\beta}(x)=e^{-i\lambda Q} h^{\alpha\beta}(x) 
e^{+i\lambda Q},\\& & \\
Q=\int\limits_{x^{0} = t}d^{3}x\, h^{\alpha\beta}(x)_{,\beta} {\stackrel{\leftrightarrow}{ \partial_{0}^{x} }} u_{\alpha}(x),
\end{array}
\label{eq:6}
\end{equation}
which leads to the infinitesimal gauge variation of the asymptotic free
graviton field
\begin{equation}
d_{Q} h^{\alpha\beta}(x)=\left[ Q,h^{\alpha\beta}(x)\right]=
-ib^{\alpha\beta\rho\sigma}u_{\rho}(x)_{,\sigma}
\label{eq:7}
\end{equation}
where $u^{\alpha}$ are c-number fields satisfying $\Box u^{\alpha}(x)=0$.
S-matrix gauge invariance: $
\lim_{g \uparrow 1}\bigl( S'(g)-S(g)\bigr)=\lim_{g \uparrow 1}\left(
-i\lambda \left[ Q,S(g)\right]+\hbox{higher com.}\right)=0$
is reached, if we can show that the `perturbative quantum gauge
 invariance '
condition~\cite{du1,is}
\begin{equation}
d_{Q}T_{n}(x_{1},\ldots , x_{n})=\left[ Q,T_{n}(x_{1},\ldots ,
 x_{n})\right]
=\hbox{divergence} 
\label{eq:8}
\end{equation}
holds true for all $ n \ge 1$. Already for $n=1$, Eq. ~(\ref{eq:8}) is non-trivial,
because $d_{Q}T_{1}^{h}(x)\neq div $; this requires the introduction
 of ghost and antighost fields, $u^{\alpha}$ and $\tilde{u}^{\beta}$,
 coupled to the graviton field through the ghost coupling~\cite{is}
\begin{equation}
T_{1}^{u}=i\kappa\left(:\tilde{u}_{\nu}(x)_{,\mu} h^{\mu\nu}(x)_{,\rho} u^{\rho}(x):
+\ldots\right),
\label{eq:9}
\end{equation}
and quantized as free fermionic vector fields $\left\{ u^{\mu}(x),\tilde{u}^{\nu}
(y)\right\}=i\eta^{\mu\nu} D_{0}(x-y)$ with infinitesimal gauge variations
$d_{Q}u^{\mu}(x)=0$ and $d_{Q}\tilde{u}^{\nu}(x)=ih^{\nu\sigma}
(x)_{,\sigma}$, so that we obtain
\begin{equation}
d_{Q}\left(T_{1}^{h}(x)+T_{1}^{u}(x)\right)=
\partial_{\nu}^{x}T_{1/1}^{\nu}(x)=\hbox{ divergence.}
\label{eq:10}
\end{equation}
The fermionic quantization is also necessary to have $Q$ nilpotent, $Q^2=0$.
The ghost fields, usually called Faddeev-Popov ghosts, are introduced
in the causal construction as a consequence of perturbative gauge invariance
Eq.~(\ref{eq:8}) for $n=1$. In the path-integral framework, the ghost fields appears as a `by-product' of the quantization after gauge fixing, but it was already noticed
by Feynman~\cite{feyn} that without ghost fields a unitarity breakdown
 occurs in 2nd
order at the loop level.

\section{Pure Quantum Gravity in 2nd Order}\label{sec:causal}

We now investigate the graviton self-energy contribution (graviton and ghost
loops) in 2nd order. The inductive construction of $T_{2}(x_{1},x_{2})$ can
be accomplished in two steps: first we construct the following causal
distribution from Eq.~(\ref{eq:4}), (\ref{eq:5}) and (\ref{eq:9}) and apply
Wick expansion
\begin{equation}
D_{2}^{\scriptscriptstyle SE}(x_{1},x_{2})=\bigl[ T_{1}^{h+u}(x_{1}),
 T_{1}^{h+u}(x_{2})\bigr]\bigr|_{\scriptscriptstyle SE}=
:h^{\alpha\beta}(x_{1})h^{\mu\nu}(x_{2}):d_{2}(x_{1}-x_{2})_
{\alpha\beta\mu\nu},
\label{eq:11}
\end{equation}
because of translation invariance the c-number distribution $d_{2}$ depends
only on the relative coordinate $x=x_{1}-x_{2}$. In momentum space we get
for the self-energy tensor-valued distribution
\begin{equation}
\hat{d}_{2}(p)_{\alpha\beta\mu\nu}=\hat{P}^{\scriptscriptstyle(4)}(p)_
{\alpha\beta\mu\nu}\Theta(
p^2)\mathop{\rm sgn}(p^{0})
\label{eq:12}
\end{equation}
where $\hat{P}^{\scriptscriptstyle (4)}(p)_{\alpha\beta\mu\nu}$ is a 
covariant polynomial of
degree 4. Then, in order to obtain $T_{2}(x)$, we split $d_{2}(x)$,
which has causal support, $\hbox{supp}(d_{2}(x))\subseteq V^{+}(x)\cup V^{-}(x)$,
into a retarded and an advanced part. This splitting must be accomplished according to the correct singular order~\cite{scha1} $\omega(d_{2})$ 
which shows
intuitively the behaviour of $d_{2}$ near the coincidence point $x=0\,$ or,
in momentum space, the UV behaviour. In this case we find $\omega(d_{2})=4$.
Thus, admitting free normalization terms with coefficients $c_{0}$, $c_{2}$
and $c_{4}$, we obtain~\cite{du1}
\begin{equation}
\hat{t}_{2}(p)_{\alpha\beta\mu\nu}=\frac{i}{2\pi}\frac{\hat{P}^{
\scriptscriptstyle (4)}(p)_{\alpha\beta\mu\nu}}{p^4}\left\{p^4 \log\left(\frac{-(p^2+i0)}{M^2}\right)+c_{0}+c_{2}p^2+c_{4}p^4\right\}.
\label{eq:13}
\end{equation}
Since $c_{4}$ can be absorbed into $M^2$ and mass and coupling constant
normalizations fix unambiguously $c_{0}=c_{2}=0$, we are left with the new parameter $M$
which defines a mass scale in the theory. We emphasize the fact that, in
virtue of the causal splitting prescription, all expressions are finite and
Eq.~(\ref{eq:13}) agrees exactly with the finite part obtained using ad-hoc
regularization schemes~\cite{caza}. As a consequence it is not necessary
 to add counterterms~\cite{hove}
to renormalize the theory. Besides, Eq.~(\ref{eq:13})
satisfies the Slavnov-Ward identity for the 2-points connected Green function~\cite{caza}:
\begin{equation}
p^{\alpha}p^{\mu}\bigl\{b_{\alpha\beta\gamma\delta}\hat{t}_{2}(p)^{\gamma
\delta\rho\sigma}b_{\rho\sigma\mu\nu}\bigr\} =0,
\label{eq:14}
\end{equation}
as well as perturbative gauge invariance Eq.~(\ref{eq:8}): $d_{Q}T_{2}^{\scriptscriptstyle SE}
(x_{1},x_{2})=div$.
For the tree graphs we quote briefly the result of Schorn~\cite{is}: perturbative gauge invariance in 2nd order, Eq.~(\ref{eq:8}), `generates'
 the 4-graviton
couplings through local normalization terms $N_{2}(x_{1},x_{2})$ of tree
graphs, in agreement with the expansion of the Hilbert-Einstein Lagrangian Eq.~(\ref{eq:3}):
$N_{2}(x_{1},x_{2})=i\kappa ^2 :{\mathcal{L}}_{\scriptscriptstyle HE}^{\scriptscriptstyle (2)}(x_{1}): \delta (x_{1}-x_{2})\,$.
For the sake of completeness, we discuss also the vacuum graphs in 2nd
 order:
in the causal perturbation theory they cannot be `divided away'
as in the GML series for connected Green functions, but this is not
 problematic
since they are finite. The corresponding $T_{2}$ distribution has singular
order $\omega =6$ and reads $\hat{T}_{2}(p)=i p^6\log\left(-(p^2+i0)/M^2 \right)+\sum_{i=0}^{3} c_{2i}(p^2)^{i}$. It is possible to show that the adiabatic (infrared) limit of vacuum graphs exists~\cite{scha2}:
$\lim_{g\uparrow 1}\bigl(\Omega ,S_{2}(g)\Omega\bigr)=0$, where $\Omega$
is the Fock vacuum of free asymptotic fields, as a consequence of the
`bad' UV behaviour of QG and, at the same time, free vacuum stability
 forces the free normalization constants to vanish.

\section{Outlook}

An interesting feature is the explicit construction of the physical Hilbert 
space of QG. In order to decouple the ghosts and the unphysical
degrees of freedom of the graviton from the theory, we could apply the
Gupta-Bleuler~\cite{gup} formalism with indefinite metric, but we prefere to realize
the free fields representations on a Fock space $\mathcal{F}$ with positive definite metric. Lorentz covariance requires then the introduction of
 the Krein
structure~\cite{kra} in $\mathcal{F}$ and we can characterize the physical subspace
${\mathcal{F}}_{phys}$ by
\begin{equation}
{\mathcal{F}}_{phys}=\ker\left\{Q,Q^{\dagger}\right\}\cap\left\{\Phi\in
{\mathcal{F}}\mid \eta_{\alpha\beta} h^{\alpha\beta}(x)^{\scriptscriptstyle (+)}\Phi =0\right\}.
\label{eq:15}
\end{equation}

\section*{Acknowledgments}

The author would like to thank the organizers of the workshop
`Quantum Field Theory under the Influence of External Conditions',
Leipzig, 14-18 September 1998; Prof. G. Scharf and the members
of the Institute for Theoretical Physics at the Z\"urich University.
This work was partially supported by the Swiss National Science
Foundation.

%\section*{References}

\end{document}